\documentclass[aps,prl,twocolumn,showpacs,superscriptaddress,groupedaddress]{revtex4}  
\usepackage{graphicx,comment,float,braket,textcomp,amsmath,lipsum,tikz}  
\usepackage{dcolumn}   
\usepackage{bm}        
\usepackage{amssymb}   
\usepackage{upgreek}

\usepackage{bbm}

\definecolor{mauve}{rgb}{0.5,0.5,0.9}
\definecolor{bblue}{rgb}{0,0.8,1}

\begin{document}

\title{Measuring Exciton Fine-Structure in Perovskite Nanocrystal Ensembles}

\author{Albert Liu}
\email{albert.liu@mpsd.mpg.de}
\affiliation{Max Planck Institute for the Structure and Dynamics of Matter, Hamburg, Germany}

\vskip 0.25cm


\begin{abstract}
    Lead-halide perovskite nanocrystals (PNCs) exhibit unique optoelectronic properties, many of which originate from a purported bright-triplet exciton fine-structure. A major impediment to measuring this fine-structure is inhomogeneous spectral broadening, which has limited most experimental studies to single-nanocrystal spectroscopies. It is shown here that the linearly-polarized single-particle selection rules in PNCs are preserved in nonlinear spectroscopies of randomly-oriented ensembles. Simulations incorporating rotational-averaging demonstrate that techniques such as transient absorption and two-dimensional coherent spectroscopy are capable of resolving exciton fine-structure in PNCs, even in the presence of inhomogeneous broadening and orientation disorder.
\end{abstract}

\maketitle

Semiconductor colloidal nanocrystals, also called colloidal quantum dots, comprise a material platform that has sustained research interest over decades due to their numerous potential applications. Indeed, their advantageous optical properties have found use in many areas such as displays \cite{Choi2018}, photovoltaics \cite{Carey2015}, and biological tagging \cite{Martynenko2017}.

In recent years, successful synthesis of lead-halide perovskite nanocrystals (PNCs) \cite{Protesescu2015} has spurred a renaissance in colloidal nanocrystal research. Many properties of PNCs, such as their high defect tolerance \cite{Huang2017} and unusual brightness, defy our traditional understanding of colloidal nanocrystals. In particular, recent evidence suggests a bright-triplet exciton ground state as the underlying mechanism for their efficient light-emission \cite{Becker2018}. Consisting of three non-degenerate states with orthogonal dipole moments \cite{Becker2018,Sercel2019}, this bright-triplet state has also exhibited promising quantum-coherent properties such as coherent single-photon emission \cite{Utzat2019} and superfluorescence \cite{Raino2018}. However, the existence of bright-triplet excitons in PNCs is far from conclusive. Indeed, a recent report has argued that the lowest-energy exciton state is in fact a dark singlet state \cite{Tamarat2019}. This controversy both highlights our poor understanding of fine-structure in PNCs and emphasizes the need for new approaches to investigating their underlying physics.

In general, homogeneous properties of colloidal nanocrystals, such as their fine-structure, are difficult to study primarily due to inhomogeneous spectral broadening in ensembles \cite{Liu2019_JCP,Liu2021_ACSNano}. Exciton fine-structure of PNCs has, thus far, primarily been investigated via single-nanocrystal linear fluorescence spectroscopy to circumvent inhomogeneity \cite{Fernee2014}. This presents two severe limitations. First, fluorescence measurements reveal radiative recombination timescales but are not sensitive to other dynamics such as intraband relaxation and coherence dephasing. Second, single-nanocrystal experiments preclude the measurement of {\it ensemble-averaged} properties that are relevant for practical applications. These reasons thus motivate the application of nonlinear spectroscopic techniques to PNC ensembles towards studying their fine-structure.

An important question concerning the study of PNC ensembles is how the single-particle selection rules, defined in the nanocrystal reference frame, translate to the experimental reference frame of a randomly-oriented colloidal suspension. Indeed, proper comparison between theoretical calculations and experimental spectra require rotational-averaging of a system's optical response, which is often neglected in analysis of ensemble spectra. Here, we evaluate the effect of rotational-averaging on linearly-polarized dipole moments of bright-triplet excitons in PNCs. First, the general theoretical framework of rotational-averaging a perturbative optical-response is described. The rotational-averaged linear and third-order optical responses are simulated, in particular for transient absorption and two-dimensional coherent spectroscopy, which reveal that the linearly-polarized single-particle selection rules are preserved in nonlinear spectroscopies with appropriate choice of excitation polarization.

\section{Rotational-Averaging of Optical Responses}

In general, spectroscopy via applied electromagnetic fields may be interpreted as measurement of a material's optical response function. In the interest of describing different spectroscopic techniques under a unified framework, we express optical response functions perturbatively (valid in the limit of weak excitation field). For example, induced linear and third-order polarizations may be written:
\begin{align}
    P^{(1)}(t) &= \int^t_{-\infty}S^{(1)}(t_1)E(t-t_1)dt_1\\
    \nonumber P^{(3)}(t) &= \int^t_{-\infty}\int^{t_3}_{-\infty}\int^{t_2}_{-\infty}S^{(3)}(t_3,t_2,t_1)\\
    &\hspace{1.5cm} E(t-t_3)E(t_3-t_2)E(t_2-t_1)dt_1dt_2dt_3
\end{align}
where $S^{(1)}(t_1)$ and $S^{(3)}(t_3,t_2,t_1)$ are the linear and third-order optical response functions respectively. Here we will assume the impulsive limit (often valid in condensed systems at cryogenic temperatures), in which $P^{(n)}$ becomes identical to $S^{(n)}$ with the appropriate time arguments. Functionally, these optical response functions are constructed from quantum pathways that represent sequences of changes in the system density matrix, where the number of changes corresponds to the perturbative order of a spectroscopy and the probed response function:
\begin{align}
    \nonumber &S^{(n)}(t_n,t_{n-1},\dots t_1)\\ 
    &\hspace{1.5cm}= \sum_i\Braket{C}_i\left[R^{(n)}_i(t_n,t_{n-1},\dots t_1) + \text{c.c.}\right]
\end{align}
where $\Braket{C}_i$ are rotational-averaging coefficients \cite{Scholes2004} which scale each respective quantum pathway $R^{(n)}_i$. The value of $\Braket{C}_i$ depends on the experimental polarization configuration and the transition selection rules of each quantum pathway, specifically:
\begin{align} \label{C-Equation}
    \Braket{C} &= \sum\limits_{i_1\dots i_n,\lambda_1\dots \lambda_n}A_{i_1\dots i_n}I^{(n)}_{i_1\dots i_n,\lambda_1\dots \lambda_n}P_{\lambda_1\dots\lambda_n}
\end{align}
where $A_{i_1\dots i_n}$ and $P_{\lambda_1\dots \lambda_n}$ are tensor products of the excitation polarization and transition dipole moment vectors respectively. $I^{(n)}_{i_1\dots i_n,\lambda_1\dots \lambda_n}$ is then an interface tensor \cite{Andrews1977} which rotational-averages the transition strength tensor $P_{\lambda_1\dots \lambda_n}$ (defined in the nanocrystal reference frame $\lambda_m$) according to the polarization tensor $A_{i_1\dots i_n}$ (defined in the experimental reference frame $i_m$). Details of the rotational-averaging theory and definitions of $I^{(n)}$ are given in the Supplementary Information. We now apply the above procedure to a concrete system, the bright-triplet fine-structure in PNCs.

\section{Fine-Structure in Perovskite Nanocrystals}


The relative energies of fine-structure states result from a hierarchy of corrections to the simple spherical quantum well model of nanocrystals. In nanocrystals of inorganic semiconductors, strong electron-hole exchange interaction usually results in a optically-active spin-triplet positioned energetically above a ``dark" spin-singlet ground state. In lead-halide PNCs however, the Rashba effect \cite{Rashba1988} has been shown to play an additional significant role in excitonic fine-structure \cite{Isarov2017,Sercel2019} and may result in an optically-active bright-triplet ground state \cite{Becker2018}.

\begin{figure}[t]
    \centering
    \includegraphics[width=0.5\textwidth]{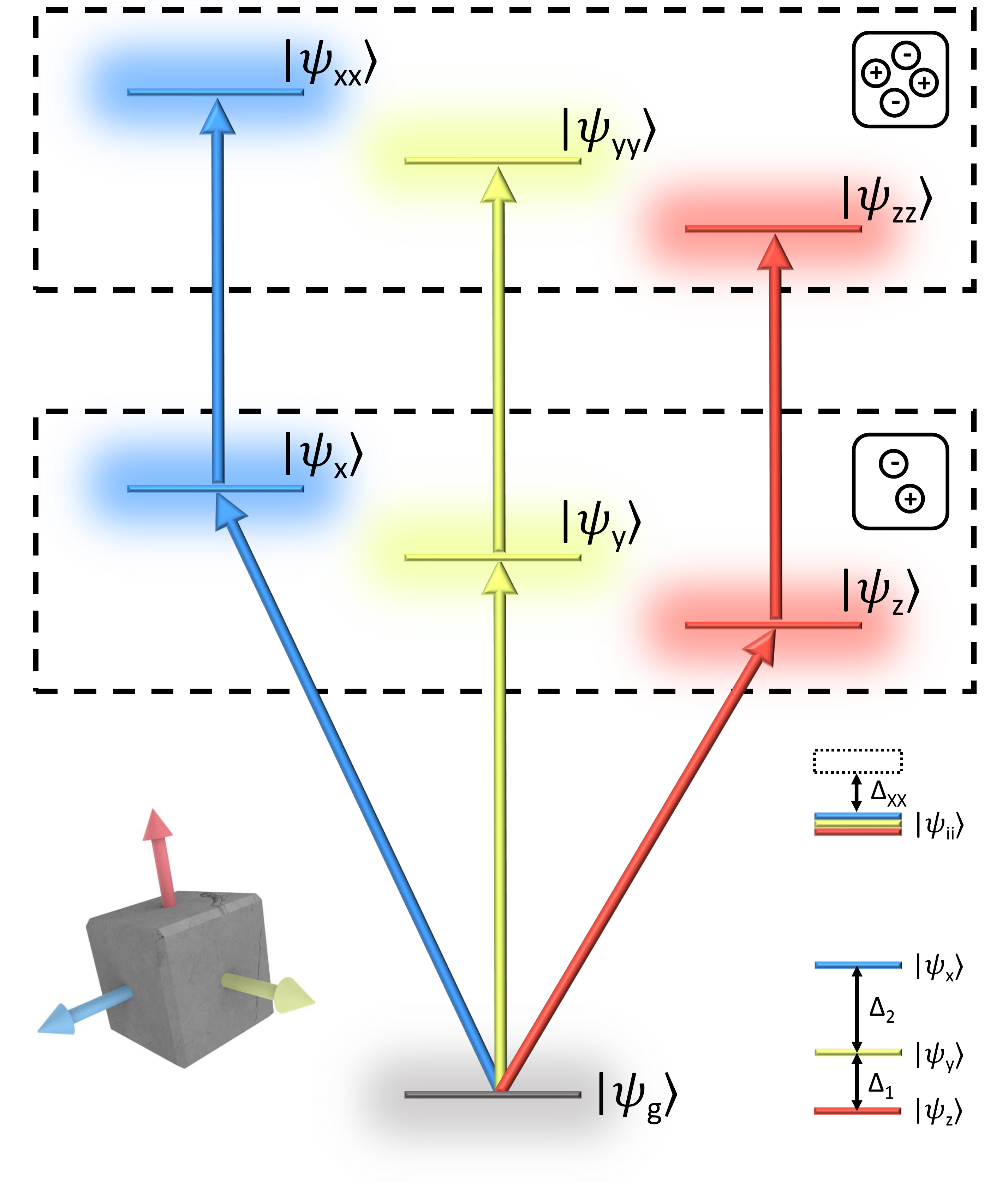}
    \caption{Fine-structure of orthorhombic phase perovskite nanocrystals. Transitions from the excitonic ground state $\Ket{\psi_g}$ to the bright triplet states $\Ket{\psi_i}$, where $i = \{x,y,z\}$, are polarized along the orthorhombic symmetry axes. Doubly-excited biexciton states $\Ket{\psi_{ii}}$ are also shown, which inherit the respective selection rules of singly-excited transitions \cite{Yin2017}. We note that the case of a fully non-degenerate triplet manifold shown here assumes an orthorhombic phase perovskite lattice \cite{Tamarat2019}.}
    \label{Fig1}
\end{figure}

\begin{table}[t]
\caption{\label{Table1} Rotational-averaging coefficients $\Braket{C}$ of third-order quantum pathways beginning with excitation of $\Ket{\psi_x}$ for each polarization scheme. Remaining coefficients may be determined by cyclic permutation of the dipole moments $\{{\bm \mu}_{\text X},{\bm \mu}_{\text Y},{\bm \mu}_{\text Z}\}$.}
\begin{ruledtabular}
\begin{tabular}{c | c | c }
$A_{i_1i_2i_3i_4}$ & $P_{\lambda_1\lambda_2\lambda_3\lambda_4}$ & $\Braket{C}$ \\ \hline

${\bf p}_{\text X} \otimes {\bf p}_{\text X} \otimes {\bf p}_{\text X} \otimes {\bf p}_{\text X}$ & ${\bm \mu}_{\text X} \otimes {\bm \mu}_{\text X} \otimes {\bm \mu}_{\text X} \otimes {\bm \mu}_{\text X}$ & $\frac{1}{5}$ \\

${\bf p}_{\text X} \otimes {\bf p}_{\text X} \otimes {\bf p}_{\text X} \otimes {\bf p}_{\text X}$ & ${\bm \mu}_{\text X} \otimes {\bm \mu}_{\text Y} \otimes {\bm \mu}_{\text X} \otimes {\bm \mu}_{\text Y}$ & $\frac{1}{15}$ \\

${\bf p}_{\text X} \otimes {\bf p}_{\text X} \otimes {\bf p}_{\text X} \otimes {\bf p}_{\text X}$ & ${\bm \mu}_{\text X} \otimes {\bm \mu}_{\text Z} \otimes {\bm \mu}_{\text X} \otimes {\bm \mu}_{\text Z}$ & $\frac{1}{15}$ \\

${\bf p}_{\text X} \otimes {\bf p}_{\text X} \otimes {\bf p}_{\text X} \otimes {\bf p}_{\text X}$ & ${\bm \mu}_{\text X} \otimes {\bm \mu}_{\text X} \otimes {\bm \mu}_{\text Y} \otimes {\bm \mu}_{\text Y}$ & $\frac{1}{15}$ \\

${\bf p}_{\text X} \otimes {\bf p}_{\text X} \otimes {\bf p}_{\text X} \otimes {\bf p}_{\text X}$ & ${\bm \mu}_{\text X} \otimes {\bm \mu}_{\text X} \otimes {\bm \mu}_{\text Z} \otimes {\bm \mu}_{\text Z}$ & $\frac{1}{15}$ \\ \hline

${\bf p}_{\text X} \otimes {\bf p}_{\text X} \otimes {\bf p}_{\text Y} \otimes {\bf p}_{\text Y}$ & ${\bm \mu}_{\text X} \otimes {\bm \mu}_{\text X} \otimes {\bm \mu}_{\text X} \otimes {\bm \mu}_{\text X}$ & $\frac{1}{15}$ \\

${\bf p}_{\text X} \otimes {\bf p}_{\text X} \otimes {\bf p}_{\text Y} \otimes {\bf p}_{\text Y}$ & ${\bm \mu}_{\text X} \otimes {\bm \mu}_{\text Y} \otimes {\bm \mu}_{\text X} \otimes {\bm \mu}_{\text Y}$ & -$\frac{1}{30}$ \\

${\bf p}_{\text X} \otimes {\bf p}_{\text X} \otimes {\bf p}_{\text Y} \otimes {\bf p}_{\text Y}$ & ${\bm \mu}_{\text X} \otimes {\bm \mu}_{\text Z} \otimes {\bm \mu}_{\text X} \otimes {\bm \mu}_{\text Z}$ & -$\frac{1}{30}$ \\

${\bf p}_{\text X} \otimes {\bf p}_{\text X} \otimes {\bf p}_{\text Y} \otimes {\bf p}_{\text Y}$ & ${\bm \mu}_{\text X} \otimes {\bm \mu}_{\text X} \otimes {\bm \mu}_{\text Y} \otimes {\bm \mu}_{\text Y}$ & $\frac{2}{15}$ \\

${\bf p}_{\text X} \otimes {\bf p}_{\text X} \otimes {\bf p}_{\text Y} \otimes {\bf p}_{\text Y}$ & ${\bm \mu}_{\text X} \otimes {\bm \mu}_{\text X} \otimes {\bm \mu}_{\text Z} \otimes {\bm \mu}_{\text Z}$ & $\frac{2}{15}$ \\ \hline

${\bf p}_{\text X} \otimes {\bf p}_{\text Y} \otimes {\bf p}_{\text X} \otimes {\bf p}_{\text Y}$ & ${\bm \mu}_{\text X} \otimes {\bm \mu}_{\text X} \otimes {\bm \mu}_{\text X} \otimes {\bm \mu}_{\text X}$ & $\frac{1}{15}$ \\

${\bf p}_{\text X} \otimes {\bf p}_{\text Y} \otimes {\bf p}_{\text X} \otimes {\bf p}_{\text Y}$ & ${\bm \mu}_{\text X} \otimes {\bm \mu}_{\text Y} \otimes {\bm \mu}_{\text X} \otimes {\bm \mu}_{\text Y}$ & $\frac{2}{15}$ \\

${\bf p}_{\text X} \otimes {\bf p}_{\text Y} \otimes {\bf p}_{\text X} \otimes {\bf p}_{\text Y}$ & ${\bm \mu}_{\text X} \otimes {\bm \mu}_{\text Z} \otimes {\bm \mu}_{\text X} \otimes {\bm \mu}_{\text Z}$ & $\frac{2}{15}$ \\

${\bf p}_{\text X} \otimes {\bf p}_{\text Y} \otimes {\bf p}_{\text X} \otimes {\bf p}_{\text Y}$ & ${\bm \mu}_{\text X} \otimes {\bm \mu}_{\text X} \otimes {\bm \mu}_{\text Y} \otimes {\bm \mu}_{\text Y}$ & -$\frac{1}{30}$ \\

${\bf p}_{\text X} \otimes {\bf p}_{\text Y} \otimes {\bf p}_{\text X} \otimes {\bf p}_{\text Y}$ & ${\bm \mu}_{\text X} \otimes {\bm \mu}_{\text X} \otimes {\bm \mu}_{\text Z} \otimes {\bm \mu}_{\text Z}$ & -$\frac{1}{30}$

\end{tabular}
\end{ruledtabular}
\end{table}

An energy-level diagram of the bright-triplet ground state is shown in Fig.~\ref{Fig1}. The triplet states are denoted $\{\Ket{\psi_x},\Ket{\psi_y},\Ket{\psi_z}\}$ according to their dipole moments polarized along the lattice symmetry-axes \cite{Becker2018,Sercel2019}. Biexciton formation is also permitted between identically-polarized triplet states \cite{Yin2017}, which are denoted $\{\Ket{\psi_{xx}},\Ket{\psi_{yy}},\Ket{\psi_{zz}}\}$. The orthogonal linear dipole moments of the triplet states give rise to well-defined optical selection rules, which may be exploited in various spectroscopies of PNCs. The transition dipole moment vectors are then defined: 
\begin{align}
    {\bm \mu}_X = \begin{pmatrix} 1\\ 0\\ 0\end{pmatrix},\quad {\bm \mu}_Y = \begin{pmatrix} 0\\ 1\\ 0\end{pmatrix},\quad {\bm \mu}_Z = \begin{pmatrix} 0\\ 0\\ 1\end{pmatrix}
\end{align}
which, to emphasize, are defined in the nanocrystal reference frame. It was previously shown by Scholes \cite{Scholes2004} that circularly-polarized selection rules are preserved in a rotational-averaged nanocrystal ensemble, which is not entirely surprising when considering the decomposition of a circularly-polarized basis into two (orthogonal) out-of-phase linearly-polarized components that span the experimental observation plane. However, it is less obvious whether the linearly-polarized selection rules of PNCs persist upon rotational-averaging.

\section{One- and Two-Photon Absorption}

One- and two-photon absorption (which we refer to as 1PA and 2PA) are among the most common spectroscopic techniques applied to PNC ensembles. As linear and third-order spectroscopies respectively \cite{Friedrich1982}, the effects of rotational-averaging differ between 1PA and 2PA spectra quite surprisingly and serve as a starting point for our analysis. 

We consider the simple case of linearly-polarized excitation. Taking the linear-polarization axis as $\hat{x}$, the 1PA and 2PA polarization tensors become:
\begin{align}
    A^{\text{1PA}} = {\bf p}_{\text X} \otimes {\bf p}_{\text X}, \quad\quad A^{\text{2PA}} = {\bf p}_{\text X} \otimes {\bf p}_{\text X} \otimes {\bf p}_{\text X} \otimes {\bf p}_{\text X}
\end{align}
Examining the $\Ket{\psi_x}$ transition, the transition strength tensors become:
\begin{align}
    P^{\text{1PA}} = {\bm \mu}_{\text X} \otimes {\bm \mu}_{\text X}, \quad\quad P^{\text{2PA}} = {\bm \mu}_{\text X} \otimes {\bm \mu}_{\text X} \otimes {\bm \mu}_{\text X} \otimes {\bm \mu}_{\text X}
\end{align}
Note our assumption that transition moment vectors involving the virtual state mediating 2PA are identical to that of 1PA. This may not be the case for 2PA selection rules in PNCs \cite{Fedorov1996}. We may now evaluate $\Braket{C}$ using equation (\ref{C-Equation}) for both cases to find:
\begin{align}
    \Braket{C}^{\text{1PA}} = \frac{1}{3}, \quad\quad \Braket{C}^{\text{2PA}} = \frac{1}{5}
\end{align}
We thus find that 1PA and 2PA of linearly-polarized light are weakened by a factor of three and five respectively, whose coefficients differ solely due to the order of each spectroscopy. These calculations are extended to a general third-order optical response in the next section.

\begin{figure}[b]
    \centering
    \includegraphics[width=0.5\textwidth]{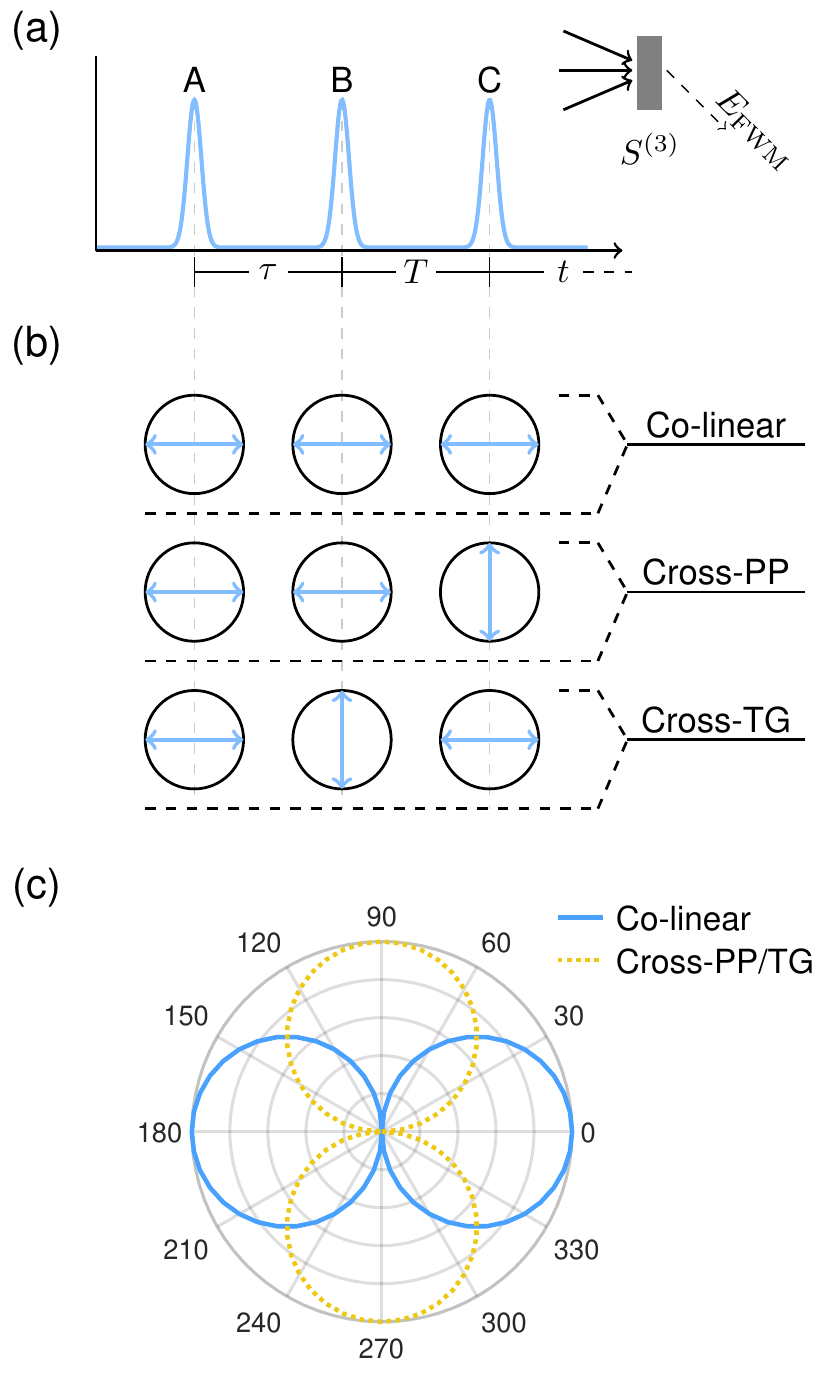}
    \caption{(a) General FWM pulse-ordering diagram. (b) Schematics of co-linear, cross-linear pump-probe (cross-PP), and cross-linear transient-grating (cross-TG) polarization schemes. Indicated polarizations correspond to pulses A, B, and C from left to right respectively. (c) Polarization dependences of the rotational-averaged FWM signals for each polarization scheme, in which the linearly-polarized selection rules are preserved.}
    \label{Fig2}
\end{figure}

\section{Four-Wave Mixing}

In generic terms, four-wave mixing (FWM) describes sets of four field-matter interactions mediated by the third-order optical response of a material. For example, 2PA is a FWM process involving four field-excitation interactions from a single applied field, and is thus proportional to the excitation intensity squared. In the most general case shown in Fig.~\ref{Fig2}a however, each field-matter interaction may arise from a distinct applied field, in which three excitation pulses $\{A,B,C\}$ generate an emitted FWM signal $E_{\text{FWM}}$ via the third-order optical response function $S^{(3)}$.

The higher dimensionality of $S^{(3)}$ affords correspondingly more numerous experimental degrees of freedom. For measurement of fine-structure in PNCs in particular, varying the polarizations of each excitation pulse yields access to a much richer set of phenomena {\it between} triplet states. We examine three polarization configurations shown in Fig.~\ref{Fig2}b, which we denote co-linear, cross-linear pump-probe (cross-PP), and cross-linear transient-grating (cross-TG). The rotational averaging coefficients for each polarization scheme are listed in Table~\ref{Table1}. We note that excitation with cross-PP and cross-TG result in a negative coefficient $\Braket{C}$ (resulting in a half-period phase shift of the emitted signal) for certain dipole transition sequences, which has also been predicted and observed for circularly-polarized selection rules \cite{Scholes2004,Scholes2006}. To answer whether the triplet-exciton selection rules persist in a randomly-oriented ensemble, we first simulate, for each polarization confguration, the total rotational-averaged FWM signal from the energy level structure plotted in Fig.~\ref{Fig1} with energy splittings $\Delta_1 = 0.8$ meV and $\Delta_2 = 2.2$ meV. Dephasing and population relaxation times of $T_2 = 25$ ps and $T_1 = 200$ ps respectively are assumed for all three triplet state transitions \cite{Becker2018_2}, and completely orthogonal dipole moments for the three triplet states (not the case in real PNCs \cite{Becker2018,Tamarat2019,Liu2021SciAdv}) are assumed for simplicity. As shown in Fig.~\ref{Fig2}c the FWM signal indeed remains polarized and varies with excitation polarization, demonstrating the feasibility of polarization-resolved FWM measurements to probe triplet-state fine-structure in PNC ensembles.

\begin{figure}[t]
    \centering
    \includegraphics[width=0.5\textwidth]{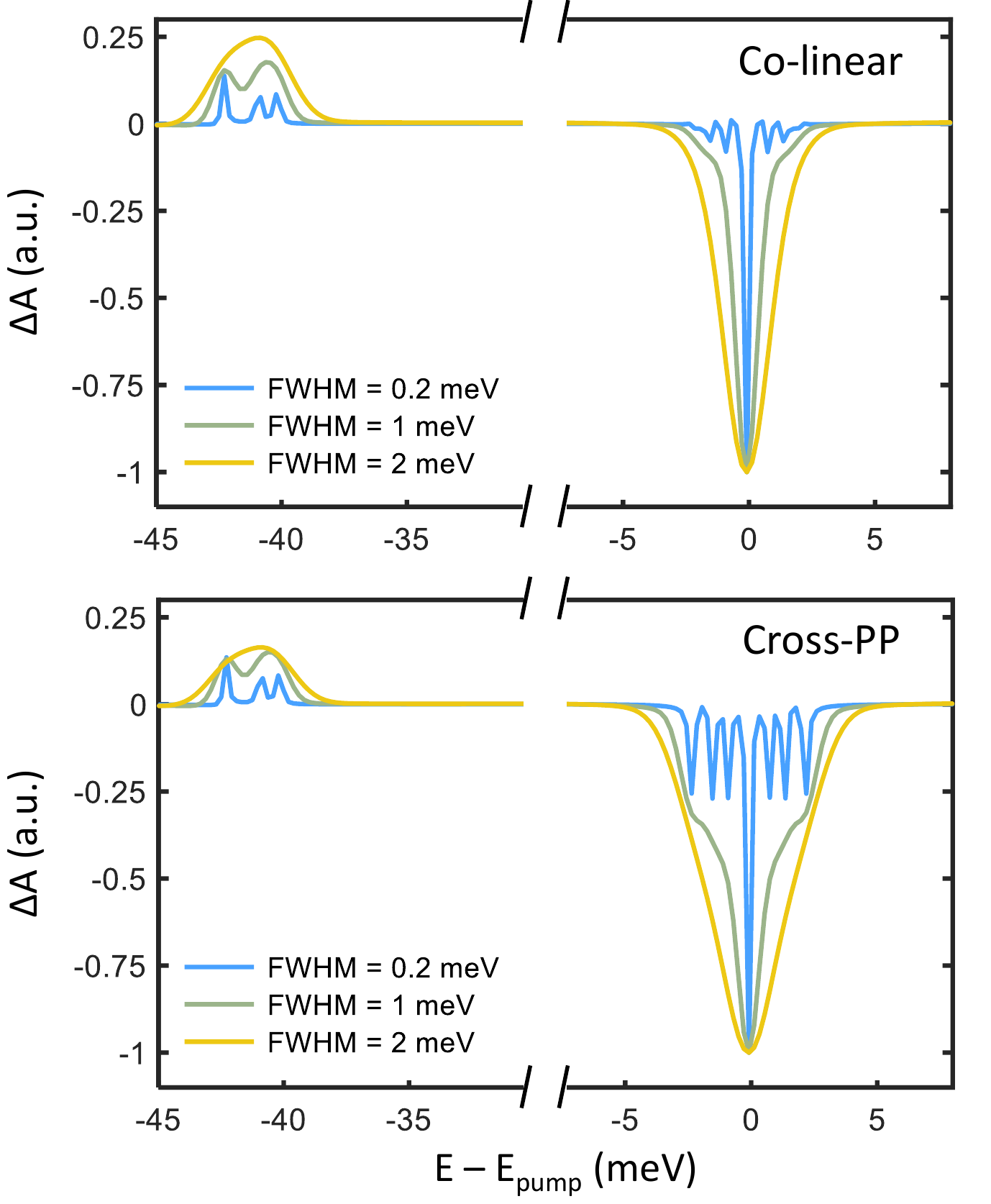}
    \caption{Transient absorption simulated for co-linear (top) and cross-linear (bottom) excitation at a pump-probe delay $T = 1$ ps. Three curves of varying pump spectral selectivity are simulated for each polarization scheme, with pump spectra centered at $E_{\text{pump}}$ = 2590 meV and FWHM of 0.2, 1, and 2 meV.}
    \label{Fig3}
\end{figure}

\subsection{Transient Absorption (Pump-Probe)}

Transient absorption spectroscopy is a spectroscopic technique commonly used to measure excited state dynamics in colloidal nanocrystals. Specifically, an initial (pump) pulse excites a material into a non-equilibrium state and the pump-induced (often spectrally-resolved) change in absorption of a subsequent (probe) pulse is measured. By performing these measurements as a function of the inter-pulse time-delay, dynamics such as intraband energy redistribution and interband relaxation may be resolved on ultrafast timescales.

Transient absorption spectroscopy may also be framed in a more general four-wave mixing perspective. For the pulse sequence in Fig.~\ref{Fig2}a, we may set $\tau = 0$ so that pulses $A$ and $B$ are jointly equivalent to the initial pump pulse and pulse $C$ acts as the probe pulse. Assuming an infinitely broad probe pulse spectrum (which approximates the usual case of a white-light continuum probe pulse), the real-quadrature of the resultant optical response function then provides the transient absorption signal \cite{Do2017,Kramer2017}:
\begin{align}
    \Delta A(\omega_t,T) \approx \text{Re}\int^\infty_{-\infty}E_{\text{pump}}(\omega_\tau)S^{(3)}(\omega_t,T,\omega_\tau)d\omega_\tau
\end{align}
where the response function $S^{(3)}(\omega_t,T,\omega_\tau)$ has been Fourier transformed along the variables $\tau$ and $t$ into frequency-space. The pump-spectrum $E_{\text{pump}}(\omega)$ then windows the optical response, enabling varying degrees of spectral selectivity.

\begin{figure*}
    \centering
    \includegraphics[width=1\textwidth]{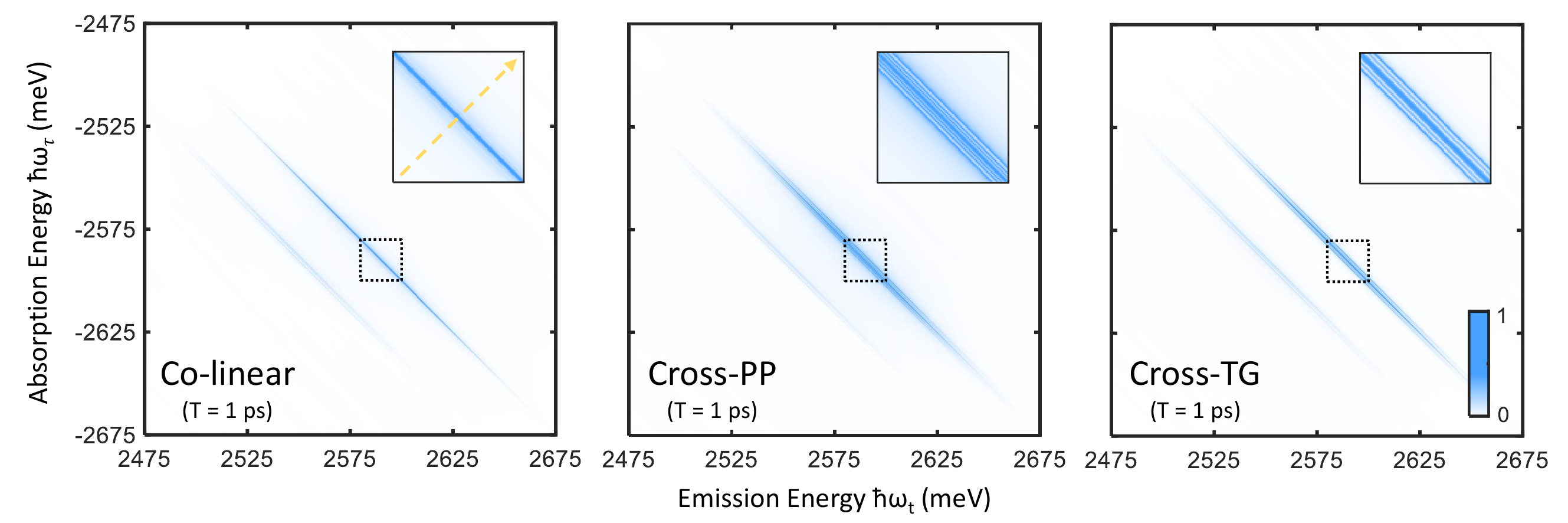}
    \caption{Rephasing one-quantum spectra simulated for co-linear (left), cross-PP (middle), and cross-TG (right) excitation and $T = 1$ ps. The sections of the spectra outlined by the dashed boxes are plotted inset, which show the appearance of sidebands with cross-PP and cross-TG excitation. The yellow dashed arrow in the left inset plot indicates the position and direction of the cross-diagonal slices shown in Fig.~\ref{Fig5}.}
    \label{Fig4}
\end{figure*}

We now simulate transient absorption spectra of PNCs by enumerating all quantum pathways comprising the third-order optical response. There are three categories of third-order quantum pathways $R^{(3)}$, named excited state emission (ESE), ground state bleach (GSB), and excited state absorption (ESA), each of which corresponds to a distinct dynamical process. In order, ESE is a negative change in absorption arising from a pump-induced excited state population which then undergoes emission stimulated by the probe pulse. GSB is also a negative absorption change due to pump-induced depletion of the absorption ground state. ESA is the only positive absorption change from a singly- to doubly-excited (biexciton) state transition. The third-order response function may be calculated by summing each ESE, GSB, and ESA quantum pathway (detailed in the Supplemental Information), and the resultant rotational-averaged transient absorption spectra are plotted in Fig.~\ref{Fig3}. For broadband excitation (FWHM = 1 and 2 meV), a broad absorption bleach feature appears at $E - E_{\text{pump}} = 0$ from the band-edge exciton transitions and a similarly broad pump-induced absorption feature appears at $E - E_{\text{pump}} = -\Delta_{\text{XX}}$ from doubly-excited state transitions. Upon narrowing the excitation bandwidth (FWHM = 0.2 meV), narrow features appear that arise from absorption and emission involving different pairs of triplet states, as well as their corresponding doubly-excited transitions. The energy of each peak then corresponds to the energy separation of the two triplet states involved.

Although transient absorption measurements with narrowband excitation can characterize triplet state energy splittings of a PNC ensemble, this method has certain drawbacks. For example, the linewidths of the fine-structure features are convolved with the pump spectral width, which interferes with extracting the homogeneous linewidth of each transition. The primary disadvantage arises from the time-bandwidth product relation for transform-limited pulses, which informs a minimum pulse duration for a given spectral bandwidth. For example, a spectral bandwidth of 0.2 meV centered at 2590 meV requires a pulse duration greater than approximately 3 picoseconds. Indeed, one may consider narrow pump-bandwidth transient absorption measurements as approaching the limit of continuous-wave spectral hole burning \cite{Liu2020}. Ultrafast dynamical processes of the triplet state manifold, such as spectral diffusion or energy transfer, thus require a more advanced spectroscopic technique to resolve.

\section{Two-Dimensional Coherent Spectroscopy}

Two-dimensional coherent spectroscopy (2DCS), or more generally multi-dimensional coherent spectroscopy \cite{Cundiff2013,Smallwood2018}, is a nonlinear spectroscopic technique capable of resolving all dimensions of a complex-valued nonlinear optical response. As the name implies, a 2-D spectrum is obtained that reflects a cross-section of $S^{(n)}$ along two spectral axes. In doing so, it offers many advantages over other one-dimensional spectroscopies that probe only an integrated cross-section of $S^{(n)}$ (such as the previously discussed transient absorption spectroscopy). 

To perform 2DCS of a third-order optical response, a transient FWM signal is measured as a function of two time variables (see Fig.~\ref{Fig2}a). As described above, this corresponds to direct measurement of $S^{(3)}(t,T,\tau)$ in the impulsive limit. By Fourier transforming $S^{(3)}$ along the two time axes, a 2-D spectrum is produced that correlates the dynamics of a system following each excitation pulse. As a Fourier transform spectroscopy, the spectral resolution of 2DCS is determined by the maximum measurement time delays (irrespective of excitation spectral bandwidth) and is therefore ideal for measuring low-energy excitations. Here we discuss two types of third-order 2-D spectra relevant to measuring fine-structure in PNCs, namely one-quantum and zero-quantum 2-D spectra.

\subsection{One-Quantum 2-D Spectra}

Rephasing one-quantum spectra are obtained by Fourier transforming a photon echo FWM signal \cite{Cundiff2012} along the time variables $\tau$ and $t$, which correlates the absorption and emission dynamics of a material (along the two conjugate axes $\omega_\tau$ and $\omega_t$ respectively) in $S(\omega_t,T,\omega_\tau)$. The rephasing nature of the photon echo FWM signal, reflected in the oppositely-signed frequencies $\omega_\tau$ and $\omega_t$, separates homogeneous and inhomogeneous broadening contributions to resonance linewidths in two-dimensional frequency space \cite{Siemens2010,Liu2021_MQT}, and is therefore ideal in studying disordered materials such as nanocrystal ensembles \cite{Liu2019_JPCL,Liu2021_ACSNano}. Though the full complex-valued spectrum contains unique information concerning many-body effects in a given system \cite{Li2006,Kasprzak2011,Martin2018}, we present only absolute value spectra here for simplicity. One-quantum spectra of perovskite nanocrystals, simulated for the three excitation polarization schemes defined in Fig.~\ref{Fig2}, are shown in Fig.~\ref{Fig4}. Two primary features are evident in the co-linear excitation spectrum, namely (1) a peak along the diagonal line ($|\hbar\omega_\tau| = |\hbar\omega_t|$) arising from resonant absorption and emission involving the bright-triplet exciton states and (2) parallel peaks centered around ($|\hbar\omega_\tau| = |\hbar\omega_t| - \Delta_\text{XX}$) involving biexciton state transitions. With cross-PP and cross-TG excitation, sidebands around the diagonal peak appear corresponding to sequential absorption and emission involving different triplet states. 

\begin{figure}[t]
    \centering
    \includegraphics[width=0.5\textwidth]{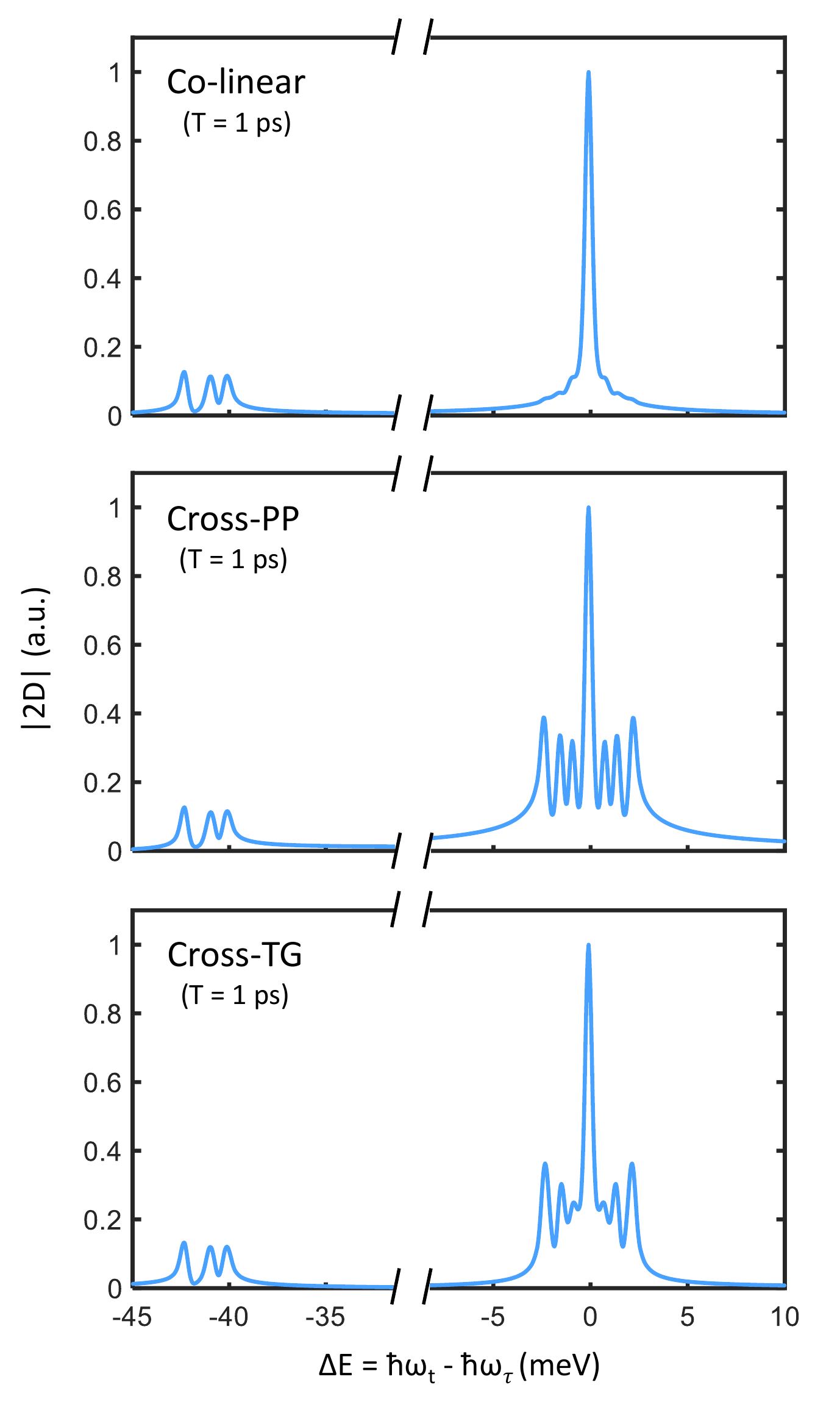}
    \caption{Transient absorption simulated for co-linear (top) and cross-linear (bottom) excitation at 5 Kelvin. Three curves of varying pump spectral selectivity are simulated for each polarization scheme, with pump spectra centered at 2590 meV and FWHM of 0.2, 1, and 2 meV.}
    \label{Fig5}
\end{figure}

We can examine these features more closely by taking slices along the direction perpendicular to the diagonal line (indicated by the dashed yellow arrow in Fig.~\ref{Fig4}). Cross-diagonal slices centered at $|\hbar\omega_t| = |\hbar\omega_\tau| = 2590$ meV are plotted in Fig.~\ref{Fig5}, which make obvious the analogy to transient absorption spectra. Just as in the narrow-excitation bandwidth transient absorption spectra in Fig.~\ref{Fig3}, coherent coupling peaks appearing at $\Delta E = \{\pm\Delta_1, \pm\Delta_2, \pm(\Delta_1 + \Delta_2)\}$ become prominent with cross-PP and cross-TG excitation. Although from summing the allowed quantum pathways one would expect identical peak amplitudes between cross-PP and cross-TG excitation, this is not observed in Fig.~\ref{Fig5}. The reason for this discrepancy is the difference in rotational-averaging coefficients as shown in Table~\ref{Table1}, which favor quantum pathways involving intermediate population states (that exhibit monotonic population relaxation during $T$) with cross-PP excitation and intermediate intraband coherences (that exhibit phase evolution during $T$) with cross-TG excitation. These simulations thus show that rephasing one-quantum spectra can resolve both energy splittings and homogeneous lineshapes of the exciton fine-structure in perovskite nanocrystals.

\subsection{Zero-Quantum 2-D Spectra}

In addition to measuring interband coherences arising from superpositions of ground and excited exciton states, 2DCS is also capable of resolving the dynamics of intraband coherences due to superpositions of two non-degenerate exciton states split by an energy within the excitation laser bandwidth. This is done by measurement of zero-quantum spectra \cite{Yang2008,Liu2019_PRL}, which are usually obtained by Fourier transforming a photon echo FWM signal along the time variables $T$ and $t$, which correlate the emission dynamics of a material with intermediate population relaxation or intraband coherence oscillations (along the two conjugate axes $\omega_t$ and $\omega_T$ respectively) in $S(\omega_t,\omega_T,\tau)$.

\begin{figure}[t]
    \centering
    \includegraphics[width=0.5\textwidth]{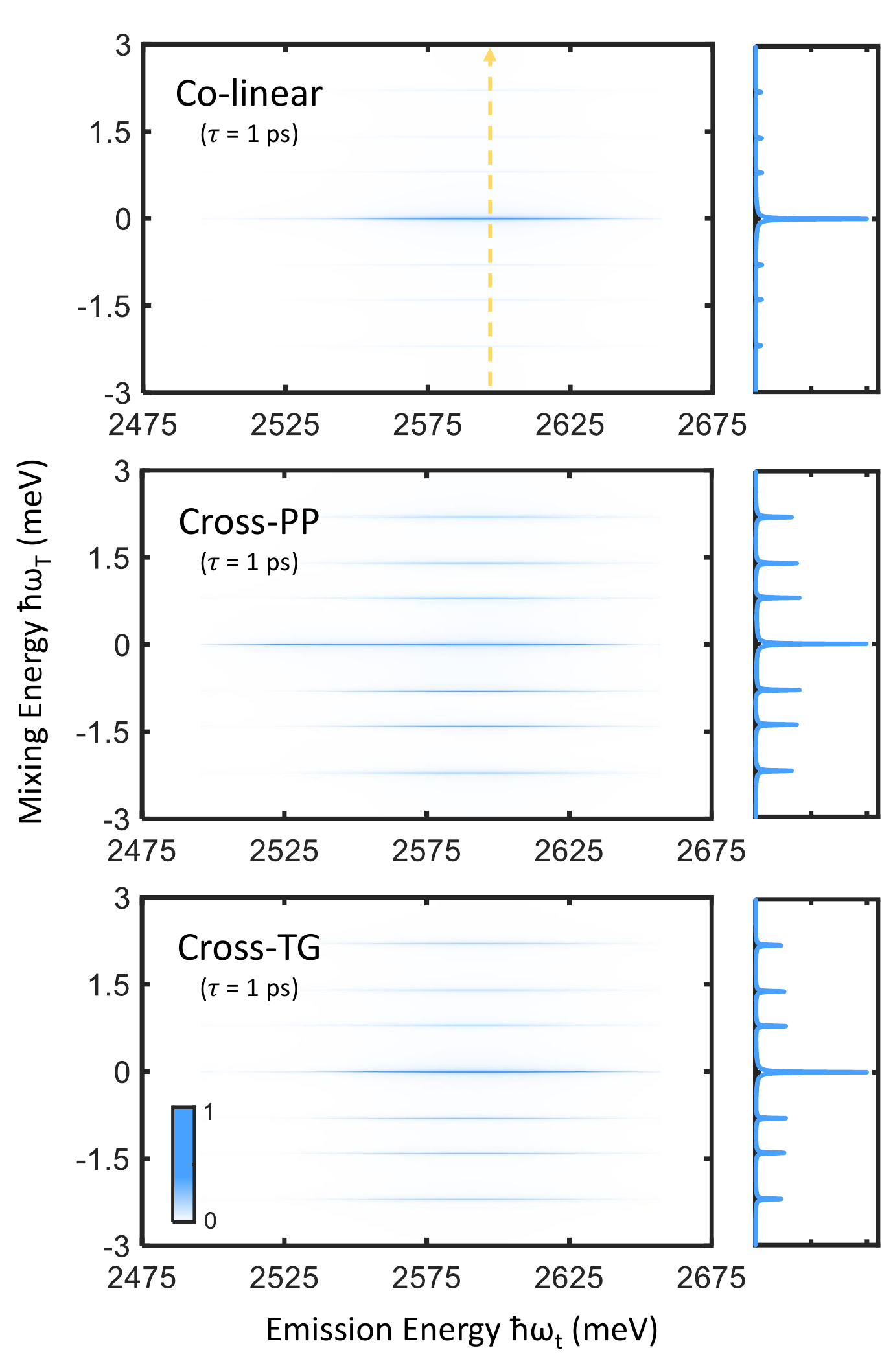}
    \caption{Zero-quantum spectra simulated for co-linear (top), cross-PP (middle), and cross-TG (right) excitation and $\tau = 1$ ps. Vertical slices taken along $\hbar\omega_t = 2590$ meV (indicated by the dashed yellow arrow) are plotted to the right of each spectrum, which emphasize the effect of excitation polarization scheme on the intraband coherence response.}
    \label{Fig6}
\end{figure}

Zero-quantum spectra of perovskite nanocrystals are shown in Fig.~\ref{Fig6}, in which primary peaks at $\hbar\omega_T = 0$ as well as sidebands at finite mixing frequency are observed. The strong central peaks may be attributed to non-oscillatory population relaxation dynamics, while the sidebands correspond to intraband coherence oscillations between pairs of bright triplet states (we note that triplet states involved in the probed intraband coherences need not be directly coupled by transition dipoles \cite{Yang2008}, and simply require a shared common ground state). Vertical slices taken along $\hbar\omega_t = 2590$ meV are also shown in Fig.~\ref{Fig6}, which shows that the intraband coherence response is enhanced by cross-PP and cross-TG excitation while being suppressed by co-linear excitation. The linewidths of the central peaks and sidebands then inform the average population relaxation rate and intraband coherence dephasing rates respectively \cite{Liu2021SciAdv}. 

\section{Conclusion}

We have presented the general framework for rotational-averaging a perturbative optical response, with applications to spectroscopy of randomly-oriented ensembles of PNCs. Via simulations of transient absorption and 2DCS spectra, we showed that the linearly-polarized single-particle selection rules of PNCs manifest in nonlinear spectroscopies of PNC ensembles with appropriate excitation polarization schemes. In particular, one-quantum and zero-quantum spectra can characterize the energy splittings and homogeneous lineshapes of the triplet state fine-structure even in the presence of both inhomogeneous broadening and orientation disorder, which has been confirmed in a recent experimental study \cite{Liu2021SciAdv}.

We emphasize that the applications of nonlinear spectroscopy to PNC ensembles extend beyond resolving the bright-triplet exciton fine-structure, the focus of the present work. For example, the directional dipole-moments of vibrational modes in lead-halide perovskites may translate to anisotropic exciton-phonon coupling, which would also manifest in a rotational-averaged optical response. Following recent work on magneto-PL spectroscopy of PNCs \cite{Tamarat2019}, magnetic-field brightening of the singlet dark state \cite{Fernee2014} may be similarly exploited in nonlinear spectroscopy of PNCs. Simultaneously resolving the homogeneous lineshapes of dark-singlet and bright-triplet states will inform the coupling mechanisms between them, and more generally the microscopic origin of exciton fine-structure in PNCs.

\end{document}